\documentclass[%
 aip,
 jmp,%
 amsmath,amssymb,
 reprint,%
unsortedaddress
]{revtex4-1}

\usepackage{graphicx}
\usepackage{dcolumn}
\usepackage{bm}
\usepackage{float}
\usepackage{gensymb}
\usepackage{color}

\begin{document}
	\title{Investigating resonance energy transfer from protein molecules to van der Waals nanosheets}

	\author{Arun Singh Patel} 
	\affiliation{School of Computational \& Integrative Sciences, Jawaharlal Nehru University, New Delhi-110067, India}
	\email{arunspatel.jnu@gmail.com}
	
	\author{Praveen Mishra} 
	\affiliation{School of Computational \& Integrative Sciences, Jawaharlal Nehru University, New Delhi-110067, India}
	
	\author{Pawan K. Kanaujia} 
		\affiliation{Nanophotonics Laboratory, Department of Physics, Indian Institute of Technology Delhi, New Delhi-110016, India}
		
	\author{Syed Shariq Husain} 
	\affiliation{School of Computational \& Integrative Sciences, Jawaharlal Nehru University, New Delhi-110067, India}
		
    \author{G.Vijaya Prakash} 
	\affiliation{Nanophotonics Laboratory, Department of Physics, Indian Institute of Technology Delhi, New Delhi-110016, India}

	\author{Anirban Chakraborti} 	
	\affiliation{School of Computational \& Integrative Sciences, Jawaharlal Nehru University, New Delhi-110067, India}
	\email{anirban@jnu.ac.in}
	
	\date{\today}

	\begin{abstract}
 The resonance energy transfer (RET) from tryptophan present in bovine serum albumin (BSA) to two dimensional (2D) nanomaterials has been reported. In these bio-nano systems, BSA molecules act as efficient donors while 2D van der Waals nanosheets of graphene, molybdenum disulfide (MoS$_2$) and tungsten disulfide (WS$_2$) act as acceptors. The fluorescence emission from tryptophan arises at 350 nm, in presence of  2D nanosheets, the fluorescence intensity of BSA  decreases. The time resolved fluorescence study shows that there is decrease in the fluorescence lifetime of BSA molecules in presence of 2D nanosheets. The decrease in fluorescence lifetime of BSA confirms the energy transfer phenomenon in bio-nano interactions. This study is important for  finding the bio-compatibility of 2D nanomaterials in biological applications.    
		
	\end{abstract}
	
	
	\maketitle
	
	In recent years, two dimensional (2D) nanosheets of van der Waals (vdW) materials have been extensively studied due to their intriguing optical, electronic, structural and chemical properties \cite{chen2013, shi2013, mak2013, falkovsky2007, ding2011}. These materials include a wide range of nanomaterials -- zero band gap to non-zero band gap. Among various types of vdW nanomaterials, graphene and transition metal dichalcogenides (TMDCs) nanosheets of the MX$_2$ type (M= Mo, W; X=S, Se, Te), are the most studied materials \cite{geim2007rise, radisavljevic2011, ramakrishna2010, wang2012}. These materials have the advantage  over other nanomaterials due to their high stability, high surface to volume ratio and a number of layer dependent electronic and optical properties \cite{hibino2009, lee2010, zhao2013}. Graphene consists of carbon atoms, arranged in a regular atomic-scale chicken wire (hexagonal) pattern. The emergence of graphene  helped further  the exploration of other two dimensional nanomaterials like, 2D sheets of TMDCs \cite{chhowalla2013, zhang2016}. Among various TMDCs nanosheets, MoS$_2$ and WS$_2$ have been lately used in various   optoelectronic devices. In bulk form, these TMDCs are found to be indirect band gap semiconductors, while for single layers, they are direct band gap semiconductors with different band gap energies \cite{anirban2016, Kin2010, Rudren2014}, e.g., MoS$_2$ has 1.2 eV indirect band gap in bulk form, while for single layer it has direct band gap of 1.8 eV \cite{shakya2016}, and similarly, WS$_2$ has 1.4 eV indirect band gap in bulk form, while in case of single layer it is a 2.1 eV direct band gap semiconductor \cite{berkdemir2013}.  
	Recently, graphene as well as TMDC nanosheets have been used in resonance energy transfer (RET) phenomenon as acceptors of energy/charge \cite{sampat2016, zang2016, kufer2016, kufer2015, goodfellow2016, Prins2014}. The RET phenomenon is observed in various physical, chemical and biological systems \cite{selvin2000, shankar2009}. In order to observe RET, one material acts as a donor and the other material as an acceptor of energy. Very recently,  Sampat et al. has studied RET from quantum dots to MoS$_2$ nanosheets \cite{sampat2016},  Goodfellow et al. used CdSe/CdS QDs as donors and WS$_2$ nanosheets as acceptor in RET \cite{goodfellow2016}. Similarly, Swathi et al. used a fluorescent dye as a donor and graphene as an acceptor for resonance energy transfer \cite{swathi2009}. There are fewer studies on  RET in biomolecules like protein and nanomaterials. Sen et al. have investigated RET in human serum albumin (HSA) protein molecules and gold nanoparticles \cite{sen2011}. Recently, Kuchlyan et al. have studied the interaction of BSA molecules with graphene oxide nanosheets \cite{kuchlyan2015}. Prior to application of nanomaterials in biological systems, testing the biocompatibility of the nanomaterials is essential. In this regard, the interaction of two dimensional nanosheets with biomolecules needs to be explored before their use in diagnosis and biomedical applications. The present study is important in the context of application of two dimensional nanosheets in bio-related fields, as for the first time here, we have investigated the bio-nano interaction in 2D nanosheets  and bovine serum albumin molecules by means of resonance energy transfer.
	 
	In our study, the vdW nanosheets were synthesized by using the liquid exfoliation method.  In a typical synthesis procedure, 50 mg of vdW powder (Sigma Aldrich) was dispersed in 10 mL of aqueous solution containing 10 mg BSA powder (Sigma Aldrich). Further the solution was ultrasonicated for 20 h using 25 W ultrasonication bath. After ultrasonication, the color of the solution changed, e.g., for MoS$_2$ from black to green, which confirmed the formation of nanosheets from bulk powder. Similarly, the color of WS$_2$ changed to brown, while that of graphene remained black. The supernatant was then collected and re-dispersed in water to prepare diluted stock solutions (0.05 mg/mL) of  vdW nanosheets. The solutions were drop casted on silicon substrates and dried at room temperature for Raman spectroscopic studies.  These nanosheets were characterized by transmission electron microscopy (TEM), Raman spectroscopy and absorption spectroscopy. The Raman spectra of graphene, MoS$_2$ and WS$_2$ nanosheets were recorded by  Renishaw inVia confocal Raman 	spectrometer using 532 nm excitation source.  For TEM study, the aqueous dispersions of the nanosheets were put on carbon coated copper grids and dried at room temperature. The TEM images of these nanosheets were recorded using  JEOL-2010 TEM setup, which was operated at an accelerating voltage of 200 kV. The absorption spectra of these nanosheets were recorded using Shimadzu (UV-2450) UV-Vis spectrophotometer.   In order to study RET in BSA and the van der Waals nanosheets, different quantities of  nanosheets were mixed with a fixed quantity of BSA containing aqueous solution. All samples were prepared  such  that  the amount of BSA molecules remains fixed  (0.5 mg in 3.5 mL solution)in all sets, while only the quantities of the nanosheets were varied in different samples. All samples were incubated  for 24 h  to achieve equilibrium.  The steady state fluorescence emissions of BSA, in absence and presence of nanosheets, were recorded  using fluorimeter with excitation wavelength of 280 nm. The time resolved fluorescence study was performed using time correlated single photon counting (TCSPC)  setup 	(FL920, Edinburgh Instruments, UK) with 284 nm light emitting diode as the excitation source. 
	
\begin{figure*}
			\centering
			\includegraphics[width=0.8\linewidth]{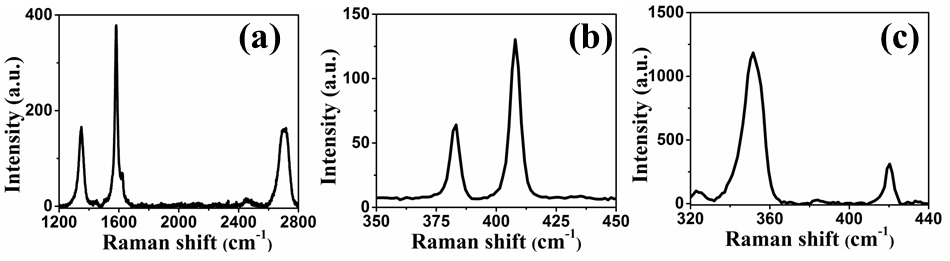} 
			\caption{Raman spectra of van der Waals nanosheets of (a) graphene, (b) MoS$_2$, and (c) WS$_2$.    \label{figure_Raman}}
		\end{figure*}

	The Raman spectra of the van der Waals nanosheets  are shown in Fig. \ref{figure_Raman}. 	
		Fig. \ref{figure_Raman}a shows the Raman spectrum of  graphene nanosheet. In case of graphene, three distinct Raman peaks are observed: 1350,  1582 and 2700 cm$^{-1}$; these peaks are known as D, G and 2D peaks, respectively \cite{Weifeng2010}. D and G peaks originate due  to	the breathing modes of sp$^2$ carbon atoms in the ring and	stretching of carbon–carbon bonds respectively. The Raman spectrum of MoS$_2$ nanosheet is shown in  Fig. \ref{figure_Raman}b. There are two distinct  Raman peaks: One at 385 cm$^{-1}$ and the other at 405 cm$^{-1}$. These peaks arise due to in-plane and out-of-plane vibrations of S-Mo-S atoms, and are known as E$_{2g}^{1}$ and A$_{1g}$ vibrational modes, respectively \cite{yang2014}.  In case of WS$_2$ nanosheets,  E$_{2g}^{1}$  and A$_{1g}$ Raman peaks are observed at 351 cm$^{-1}$ and 420 cm$^{-1}$, respectively.  

\begin{figure*}
				\centering
				\includegraphics[width=0.8\linewidth]{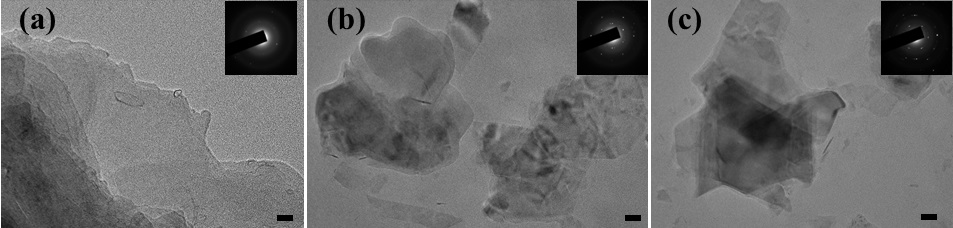} 
				\caption{TEM images of (a) graphene, (b) MoS$_2$, and (c) WS$_2$ nanosheets. Insets show selective area electron diffraction patterns of the respective 2D nanosheets.    \label{TEM}}
			\end{figure*}

	The morphologies of two dimensional vdW nanosheets of graphene, MoS$_2$ and WS$_2$, were analyzed by transmission    electron microscopy (TEM). Fig. \ref {TEM}a shows the TEM image of graphene nanosheets. The selective area electron diffraction (SAED) of graphene  is shown in the inset of Fig. \ref {TEM}a. Similarly, the TEM images of MoS$_2$ and WS$_2$ are shown in Fig. \ref {TEM}b and Fig. \ref {TEM}c, respectively, and the insets of Fig. \ref {TEM}b and Fig. \ref {TEM}c show the SAED pattern of MoS$_2$ and WS$_2$ nanosheets, respectively.  
     
\begin{figure}
			\centering
			\includegraphics[width=0.80\linewidth]{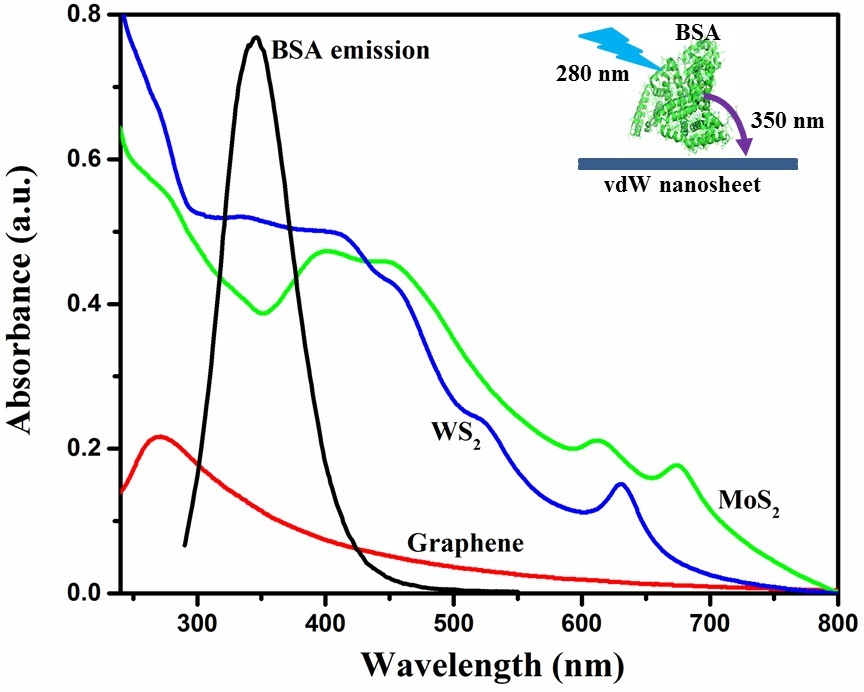} 
			\caption{Absorption spectra of graphene,   MoS$_2$ nanosheets, and  WS$_2$ nanosheets, along with the emission spectrum of BSA. Inset: Schematic representation of energy transfer between BSA and the van der Waals nanosheets.  \label{figure_abs}}
		\end{figure}

	The optical absorbance spectra of vdW nanosheets were recorded by the absorption spectrophotometer, and are shown in Fig. \ref{figure_abs}.
The absorption spectrum of graphene shows an absorption peak around 280 nm. This absorption peak arises due to  $\pi$-conjugation, which is partially restored within the  graphene sheets. In case of MoS$_2$, two absorption peaks are observed; one at 620 nm and other at 670 nm. These peaks are the resultant of transitions between valence band and conduction band of MoS$_2$.  The absorption peaks  observed at 620 nm and 670 nm are assigned as $B$ and $A$ exciton  peaks, respectively. Similarly, for WS$_2$ nanosheets the absorption peaks are observed at 530 nm and 630 nm. The absorption peaks are similar to the previous ones reported\cite{nguyen2016}.  

\begin{figure*}
			\centering
			\includegraphics[width=0.8\linewidth]{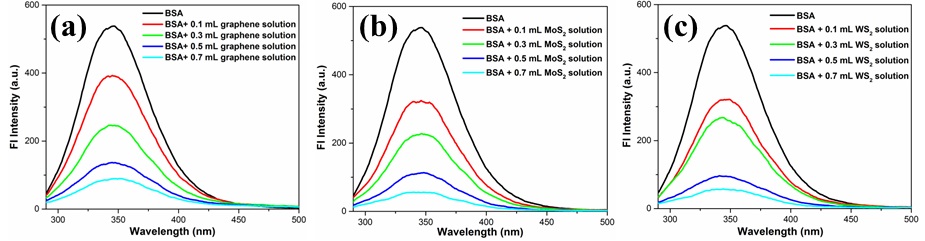} 
			\caption{Fluorescence spectra of BSA in absence and presence of  different quantities of (a) graphene, (b) MoS$_2$, and (c) WS$_2$ nanosheets, respectively.   \label{figure_BSAFl}}
		\end{figure*}

\begin{figure*}
			\centering
			\includegraphics[width=0.8\linewidth]{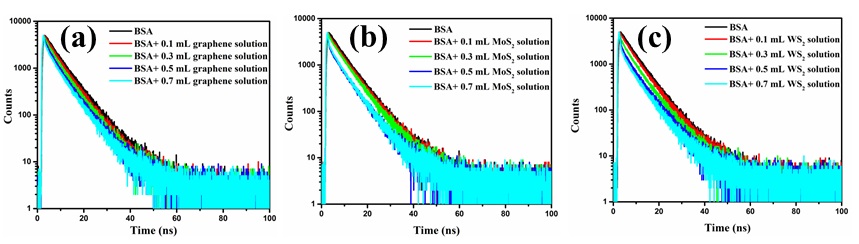} 
			\caption{Fluorescence lifetime decay curves of BSA,  in the absence or in the presence of different quantities of (a) graphene, (b) MoS$_2$, and (c) WS$_2$ nanosheets, respectively.   \label{figure_TRFS}}
		\end{figure*}

\begin{figure}
  		\centering
  		\includegraphics[width=0.95\linewidth]{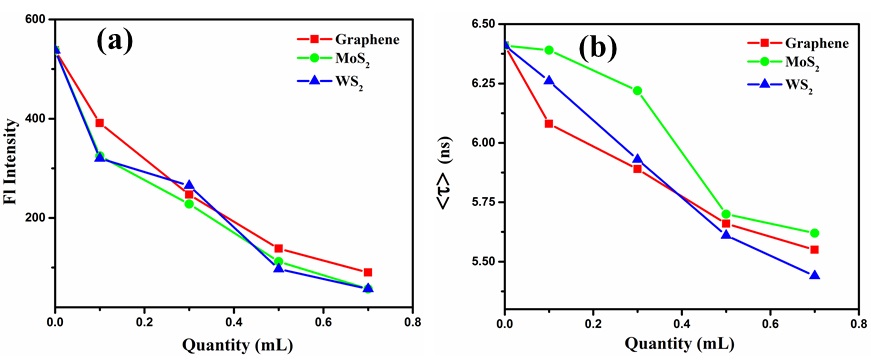} 
  		\caption{Variation of: (a) fluorescence intensity and (b) fluorescence lifetime of BSA molecules, in presence of different quantities of 2D nanosheets. \label{FlIFlT}}
  	\end{figure}

The interaction of these nanosheets with BSA molecules was studied by using fluorescence spectroscopic techniques. The steady state fluorescence spectra of BSA molecules in absence and presence of 2D nanosheets are recorded by fluorimeter using 280 nm excitation wavelength, and emission was recorded in wavelength range of 290-550 nm. The steady state fluorescence spectra of BSA molecules in presence of different concentration of 2D nanosheets  are shown in Fig. \ref{figure_BSAFl}. BSA shows fluorescence emission at 350 nm, which arises due to tryptophan present in BSA molecules \cite{tayeh2009}. 
In Fig. \ref{figure_BSAFl}a, it is found that with increase in the concentration of graphene nanosheets, the fluorescence (Fl) intensity of BSA  decreases. The decrease in Fl intensity confirms  quenching due to graphene nanosheets. In     Fig. \ref{figure_BSAFl}b  fluorescence quenching of BSA molecules in presence of  MoS$_2$ nanosheets is shown. Similarly, for WS$_2$ nanosheets, the quenching of BSA fluorescence is shown in Fig. \ref{figure_BSAFl}c. The fluorescence quenching of BSA molecules is confirmed, but  in order to ascertain the quenching nature (static or dynamic), time resolved fluorescence study was done using time correlated single photon counting technique.       
It is observed that the fluorescence lifetime of  BSA decays, in absence and presence of different concentrations of vdW nanosheets. The fluorescence lifetime decay spectra of BSA, in the absence and presence of 2D nanosheets, are shown  in Fig. \ref{figure_TRFS}. 
Fig. \ref{figure_TRFS}a shows the Fl lifetime decay curve of BSA molecules in presence of graphene nanosheets. It is observed that with increase in the quantity of graphene, the fluorescence decay curves are steeper in comparison to pure BSA. There is a systematic change in the decay curves with increase in the quantity of graphene nanosheets. Similarly, Fig. \ref{figure_TRFS}b and Fig. \ref{figure_TRFS}c show the fluorescence lifetime decays of BSA, in presence of MoS$_2$ and WS$_2$ nanosheets, respectively. These decay curves were fitted with exponential decaying functions of following form \cite{patel2014}:
 \begin{equation}
 	I(t)=\sum\limits_{j=1}^{m}\alpha_j \exp(-t/\tau_j), \label{eq_expfit}
 \end{equation}
where $m$ is the number of discrete decay components, $\tau_j$ are the decay times, and $\alpha_j$ represents the weighing parameter associated with the $j^{th}$ decay. The decay curve of  BSA in absence of nanosheets is fitted with bi-exponential ($m=2$) decaying function, while in presence of nanosheets the curves are  fitted with triple-exponential decaying function (fitted  curves not shown). The average fluorescence lifetimes of BSA in absence, as well as in presence of nanosheets, were calculated using following equation \cite{patel2016}:
   \begin{equation}
 	\langle\tau\rangle=\frac{\sum\limits_{j=1}^{m}\alpha_j\tau_j}{\sum\limits_{j=1}^{m}\alpha_j}.
\label{eq_3averagelifetime}
 \end{equation}
The average fluorescence lifetime of pure BSA was found to be 6.41 ns. In presence of different quantities of graphene, the Fl lifetime decreases -- for 0.1, 0.3, 0.5 and 0.7 mL of graphene solutions, the Fl lifetimes are found to be 6.08, 5.89, 5.66 and 5.55 ns, respectively. Thus, there is a systematic decrease in the Fl lifetimes of BSA molecules in presence of different quantities of graphene. Similarly, for MoS$_2$ nanosheets, the respective Fl lifetimes are found to be 6.39, 6.22, 5.70 and 5.62 ns in presence of 0.1, 0.3, 0.5 and 0.7 mL of MoS$_2$ nanosheets. For WS$_2$ the values are 6.26, 5.93, 5.61, 5.44 ns with 0.1, 0.3, 0.5 and 0.7 mL of WS$_2$ solutions, respectively. In all cases, with increase in the quantity of vdWs nanosheets, there is a corresponding decrease in the Fl lifetime of BSA molecules. In presence of 2D nanosheets, the fluorescence intensity as well as the fluorescence lifetime of BSA molecules are found to decrease. In  Fig \ref{FlIFlT}, variation of Fl intensity and Fl lifetime of BSA with different quantities of 2D nanosheets are shown.  The decrease in lifetime indicates that there is some energy transfer from the BSA molecules to 2D nanosheets. In Fig. \ref{figure_abs}, it is shown that the emission spectrum of BSA overlaps with the absorption spectra of vdW nanosheets. This results in transfer of some energy from BSA molecules to the nanosheets. In the process of energy transfer, the fluorophores show quenching in the Fl intensity, as well as Fl lifetimes. Using Fl lifetime data, the energy transfer efficiency from BSA molecules to 2D nanosheets has been calculated using following equation:
   \begin{equation}
   	E=1-\frac{\tau_{DA}}{\tau_D} .
   	\label{eq_energytime1}
   \end{equation}
Here, $\tau_{D}$ and $\tau_{DA}$ are fluorescence lifetimes of BSA, in absence and presence of 2D nanosheets (acceptors), respectively.  It is found that with increase in the concentration of acceptors the energy transfer efficiency increases.  
  The energy transfer efficiencies, in case of graphene are found to be 5, 8, 12 and 13 \% for 0.1, 0.3, 0.5 and 0.7 mL of graphene solutions, respectively. For MoS$_2$ the efficiencies are found to be 0.5, 3, 11 and 12 \% with 0.1, 0.3, 0.5, and 0.7 mL of MoS$_2$ solutions, respectively. Finally, in the case of WS$_2$ nanosheets, the efficiencies were 2, 7, 12 and 15 \% for 0.1, 0.3, 0.5 and 0.7 mL of the solutions, respectively. 
  
  The energy transfer efficiency was also calculated using steady state fluorescence emission data of BSA, in absence as well as presence of different vdWs nanosheets, using following equation:
     \begin{equation}
     E=1-\frac{F_{DA}}{F_D},
     \label{eq_energytime2}
     \end{equation}
where F$_D$ is fluorescence intensity of BSA in absence of acceptors, while F$_{DA}$ is intensity in presence of acceptors. The energy transfer efficiencies were found to be 27, 54, 74 and 83 \% for 0.1, 0.3, 0.5 and 0.7 mL of graphene solutions, respectively. In case of MoS$_2$, the efficiencies were 40, 58, 79 and 89 \% for 0.1, 0.3, 0.5 and 0.7 mL of MoS$_2$ solutions, respectively. Similarly, for WS$_2$, the efficiencies were found to be 40, 51, 82 and 89 \% for 0.1, 0.3, 0.5 and 0.7 mL of the solutions, respectively. 
  
  There is a difference in the energy transfer efficiency values for Fl intensity data and Fl lifetime data, which indicates possibilities of two kinds of Fl quenching mechanisms, viz., static and dynamic. The decrease in fluorescence lifetimes normally corresponds to the dynamic nature of Fl quenching, and possibility of the resonance energy transfer from BSA to 2D nanosheets, while quicker decay in the Fl intensity confirms the static nature of quenching. In Fig. \ref{figure_abs},  it has been shown that the fluorescence emission of BSA molecules overlaps with some fraction of the absorption spectrum of the 2D nanosheets. The overlapping fluorescence of BSA and absorption of nanosheets causes transfer of energy from BSA molecules to the nanosheets. When the BSA molecules are excited by 280 nm excitation light, they emit at higher wavelength of 350 nm. In presence of nanosheets, some fraction of this emission gets absorbed by the nanosheets, which results in decrease in fluorescence intensity, as well as the fluorescence lifetimes of the BSA molecules.  At the same time, the BSA molecules are attached to the nanosheets by van der Waals interaction and form BSA-nanosheets complexes, which is not fluorescent. Therefore, there is a steeper decrease in the fluorescence intensity of BSA molecules, in presence of nanosheets, as compared to decrease in the Fl lifetimes.

In summary, we have studied the resonance energy transfer between BSA molecules and two dimensional nanosheets of graphene, MoS$_2$ and WS$_2$. These nanosheets can be used as acceptors for energy transfer studied in biomolecules, which can be further explored to study the effect of 2D nanomaterials on biomolecules, prior to application of these nanomaterials in biological applications.  


AC and ASP acknowledge financial support from grant number BT/BI/03/004/2003(C) of Government of India, Ministry of Science and Technology, Department of Biotechnology, Bioinformatics division.
Authors thank the AIRF, JNU for TEM characterization and the FIST (DST, Govt. of India) UFO scheme of IIT Delhi for Raman/PL facility.

		\bibliography{main}
	


\end{document}